\documentclass[12pt]{iopart}
\usepackage{graphicx}
\begin{document}
\title[Effect of Parallel Transport Currents on the D-wave Josephson junction]
{Effect of Parallel Transport Currents on the D-wave Josephson junction}
\author{Gholamreza Rashedi}
\address{Department of Physics, University of Isfahan,
Isfahan, Iran}
\ead{rashedi@phys.ui.ac.ir}
\date{\today}
\begin{abstract}
In this paper, the nonlocal mixing of coherent current states in
$d$-wave superconducting banks is investigated. The
superconducting banks are connected via a ballistic point contact.
The banks have a mis-orientation and a phase difference.
Furthermore, they are subjected to a tangential transport current
along the $ab$-plane of $d$-wave crystals and parallel with
interface between superconductors. The effects of mis-orientation
and external transport current on the current-phase relations and
current distributions are subjects of this paper. It observed
that, at values of phase difference close to $0$, $\pi$ and $2\pi$
the current distribution may have a vortex-like form in vicinity
of the point contact. The current distribution of this junction
between $d$-wave superconductors is totally different from the
junction between $s$-wave superconductors. As the interesting
results, spontaneous and Josephson currents are observed for the
case of $\phi=0$.
\end{abstract}
 \submitto{\JPCM}
\maketitle
\section{Introduction}
The weak link between $d$-wave superconductors is a long studied
problem theoretically
\cite{YIP,BAR,TK,FSR,FOG,RAS,LOF1,LOF2,GUMANN}. A theoretical
investigation of the total transparent Josephson junction between
two $d$-wave superconductors has been done in Ref.\cite{YIP}.
Using the quasiclassical approach, a $d$-wave Josephson Junction
with low-transparent interface has been studied in Ref.\cite{BAR}.
Anisotropic and unconventional pairing symmetry has been
considered for $d-I-d$ systems and zero energy states (ZES) as the
result of sign change of the order parameter have been observed in
paper \cite{TK}. A spontaneous current parallel to the interface
between $d$-wave superconductors has been found in
paper\cite{FSR}. The junction between current-carrying states of
d-wave superconductors, has been investigated in Ref.\cite{FSR}.
Authors of paper \cite{FSR} by numerical self-consistent
calculations shows that, the supercurrent parallel to the junction
may flow in the direction opposite to the current direction at the
superconducting banks. Effect of transparency of interface of
$d$-wave Josephson junction was studied in Ref.\cite{FOG}. In
Ref.\cite{RAS}, effects of transparency and mis-orientation of two
$d-$wave crystals have been investigated analytically. Using the
Bogoliobov-De Gennes equations ZES as the origin of zero bias
conductance peak (ZBCP) were studied in Refs.\cite{LOF1,LOF2}. ZES
are introduced as the fingerprint of unconventional pairing
symmetry in\cite{LOF1,LOF2}. In Ref.\cite{GUMANN}, a special
geometry of $d-$wave superconducting layer as a weak link has been
investigated and the $\pi$ Josephson junction has been observed.
Also because of high critical temperature of $d$-wave
superconductors (and cheap production technology), many
experimental works about the $d$-wave weak link have been done in
the las two decades
\cite{GOLO,HARL,TSUEI,ILZAK1,ILZAK2,ILZAK3,TESTA}. A complete
review of these experiments has been presented in a review paper
\cite{GOLO}. A phase sensitive experiment (phase of the
superconducting order parameter) has been presented by authors of
\cite{HARL} for determination of the symmetry of order parameter
in high $T_c$ cuperate superconductors. Using the phase
interference experiments in the Josephson junctions, $d-$wave
pairing symmetry in the cuperate superconductors has been observed
in Refs.\cite{HARL,TSUEI,ILZAK1}. A nonsinusoidal form of
current-phase diagram has been observed in \cite{ILZAK2},
experimentally. Authors of Ref.\cite{ILZAK3} have measured the
current-phase relationship of symmetric grain boundary weak link
and observed that, temperature controlled sign change of the first
harmonic of the Josephson current ($I(\phi)=I_1\sin{\phi}+I_2
\sin{2\phi}+\cdot\cdot\cdot$). Because of competition between the
first and second harmonics of Josephson current, a nonmonotonic
temperature dependence of the critical current has been reported
in \cite{ILZAK3}. An experimental investigation of a Josephson
junction between $d-$wave superconductors has been done and effect
of insulator between them has been studied in \cite{TESTA}. They
observed $0-\pi$ transitions by reducing the width of insulator in
$d-I-d$ Josephson junction. From the other hand, it is well-known
that, non-locality and Josephson effect are coexisting. Charged
particles orbiting around a magnetic flux are influenced by a
magnetic flux as a phase difference although, the region including
flux is forbidden for charged particles. This phase as the
Aharonov-Bohm phase is a demonstration of non-locality of quantum
mechanics. While the supercurrent in a superconducting bulk
depends on the phase gradient locally\cite{TINK},
$\mathbf{j(r)}\propto{\nabla{\varphi{( \mathbf{r})}}}$, the
Josephson supercurrent depends on phase difference
non-locally\cite{JOS},
$\mathbf{j(\varphi_2-\varphi_1)}\propto{\sin{(\varphi_2-\varphi_1)}}\propto{(\varphi_2-\varphi_1)}$.
The interplay between local supercurrent states and non-local
phase difference between the superconducting bulks is an
interesting problem. Local supercurrent states are introduced by
superfluid velocity of Cooper pairs. The, non-locality of
Josephson current in the point contacts and the effect of
super-fluid velocity on the current states in narrow films and
wires have been studied in Ref.\cite{HEI}. An anomalous periodic
behavior in terms of magnetic flux has been observed in
Ref.\cite{HEI}. This anomalous property is demonstrated as a
result of a non-locality of supercurrent in the Josephson
junction\cite{HEI}. In addition, experimental results of
Ref.\cite{HEI} have been confirmed in analytical calculations of
Ref.\cite{LFB}. The dynamical Josephson junction between $s$-wave
superconductors has been investigated in Ref.\cite{KOLGHOL}. They
studied the quantum interference between right and left $s$-wave
superconductors, in which parallel transport currents are flowed.
The existence of two anti-symmetric vortex-like currents near the
contact and at $\phi\simeq\pi$ as a new phenomenon was reported in
Ref.\cite{KOLGHOL}. The authors of Ref.\cite{KOLGHOL} found that
the total current is not the vector sum of Josephson and the
transport currents because of a new term in the current. This term
is called the "interference" current and can be a "parallel
Josephson current". In the Ref.\cite{KOLGHOL}, effect of
reflection at the interface between $s-$wave superconductors has
been investigated analytically and numerically.

In this paper, a planar weak link between two $d$-wave
superconductors with a phase difference between order parameters
is investigated.
\begin{figure}[ht]\centering{
\resizebox{\textwidth}{0.5\textheight}{\includegraphics{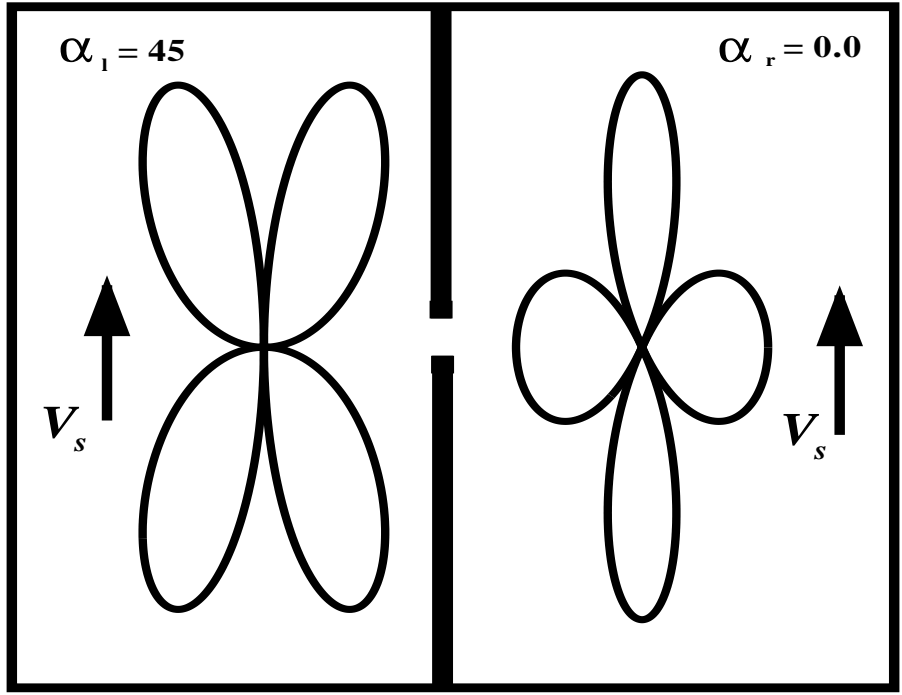}}}
 \caption{\protect\small \ Model of the contact in the insulating
partition, along the ${\bf {\hat y}}$, between two mis-oriented
$d$-wave superconducting bulks with transport supercurrent on the
banks. The plane of paper is $ab$-plane of $d$-wave
superconductors. In the $d$-wave superconductors like $YBaCuO$ the
$ab$-plane is the plane of $CuO$. \hskip0truecm} \label{fig1}
\end{figure}
 The $ab$-planes of two superconductors have a
mis-orientation and $c$ axis of two crystals are parallel to the
interface between $d$-wave superconductors. In the center of the
interface we create an ideal transparent thin slit with length $L$
and width $a$. Interference between wave functions of the left and
right superconductors occurs through the slit. The remainder part
of interface is an ideal insulator and is impenetrable for Cooper
pairs. Also, two transport supercurrents at the $ab$-planes are
flowing parallel to the insulator and the contact plane (see
Fig.\ref{fig1}). The Josephson current from one of the bulks to
another, is a result of the interference between states with phase
difference $\phi$ as is predicted in Ref.\cite{JOS}. The contact
scales (thin slit), length $L$ and width $a$, are much larger than
the Fermi wavelength and smaller than coherence length of
superconductivity. Furthermore, these scales are small as compared
with the mean free path of quasi-particles. Therefore the
quasi-classical approximation for the ballistic point contact can
be used. The Eilenberger equations for this system are solved and
Green functions are obtained. The effects of mis-orientation and
phase difference between order parameters and super-fluid velocity
on the current distributions and current-phase graphs are
investigated in this paper.

The organization of the rest of this paper is as follows. In
Sec.\ref{2} the quasiclassical equations for Green functions are
presented. The obtained formulas for the Green functions are used
to analyze a current state in the ballistic point contact. Also
the effects of transport current and mis-orientation on the
current distribution at the contact plane are investigated. In
Sec.\ref{3} the results of simulation for current distribution in
vicinity of the contact will be investigated. An analytical
investigation of system near the critical temperature will be done
in Sec.\ref{4}. The paper will be finished with some conclusions
in Sec.\ref{5}.
\section{Formalism and Basic Equations}\label{2}
 The Eilenberger equations for the
$\xi$-integrated Green's functions are used to describe the
coherent current states in a superconducting ballistic
micro-structures \cite{EIL}:
\begin{figure}[ht]\centering{
\resizebox{\textwidth}{0.5\textheight}{\includegraphics{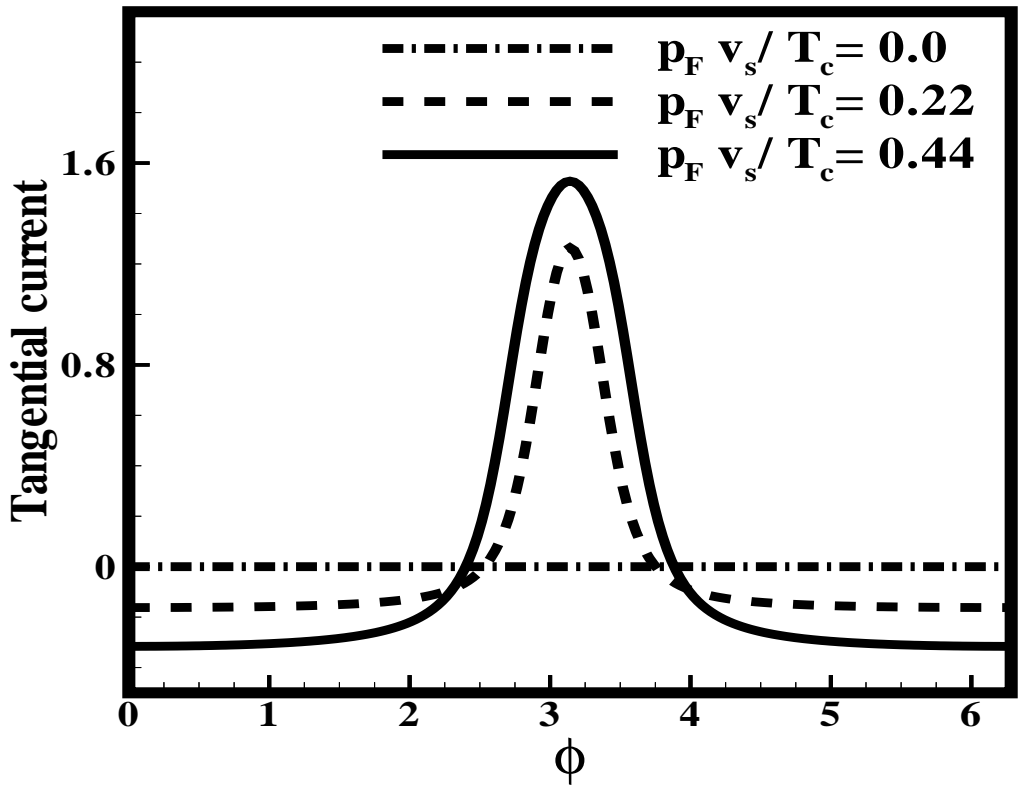}}}
\caption{\protect\small \ Tangential current $j_{y}$ versus $\phi$
for $T/T_c=0.1$, $\alpha_l=\alpha_r=0$ in the units of
$j_{0}=4\protect\pi e N(0)v_{F}T$. This is like the case of
junction between conventional superconductors\cite{KOLGHOL}
\hskip0truecm} \label{fig2}
\end{figure}
\begin{figure}[ht]\centering{
\resizebox{\textwidth}{0.5\textheight}{\includegraphics{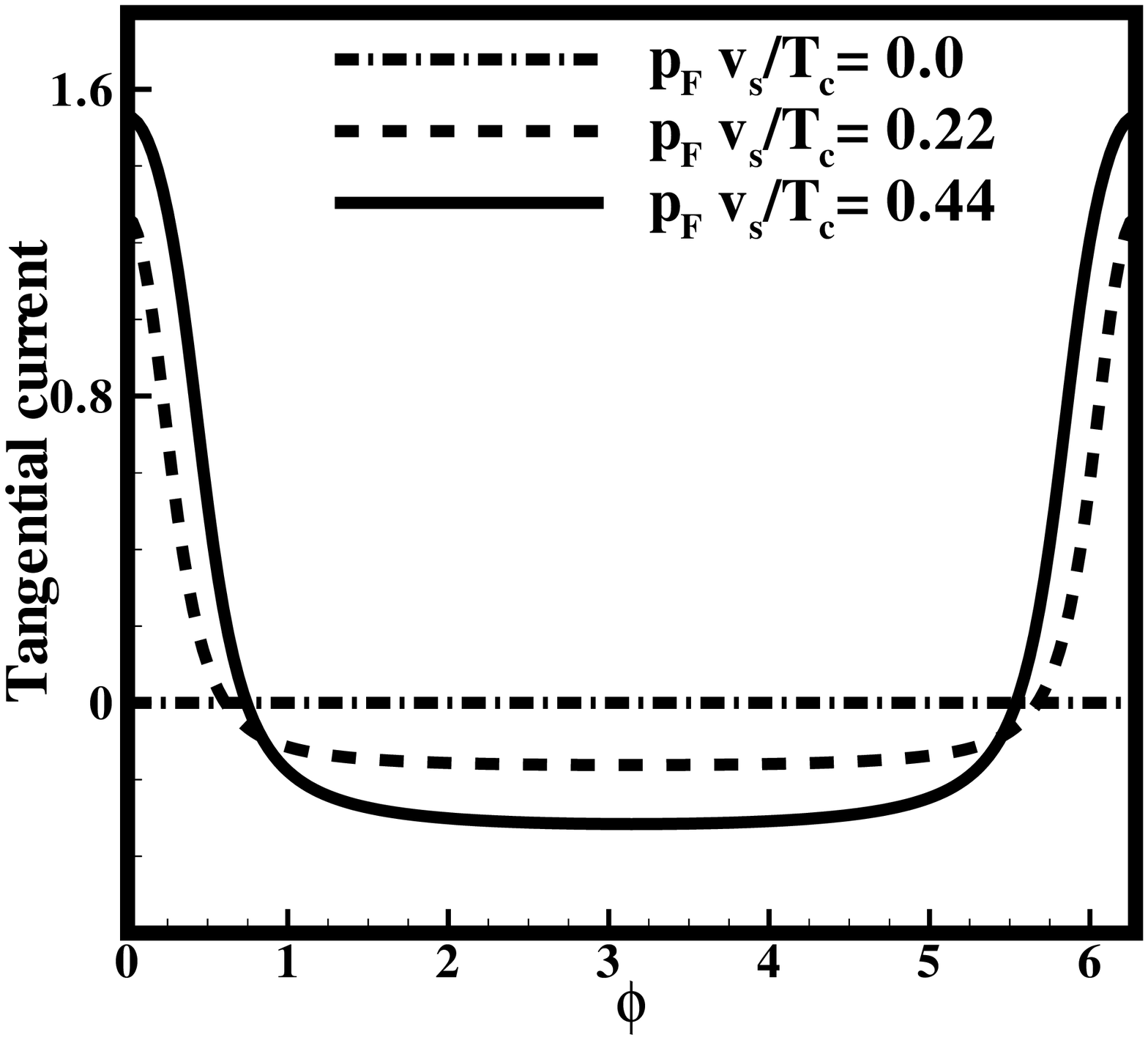}}}\caption{\protect\small
\ Tangential current $j_{y}$ versus phase $\phi$ for $T/T_c=0.1$,
$\alpha_{l}=0$ and $\alpha_{r}=\frac{\pi}{2}$. \hskip0truecm}
\label{fig5}
\end{figure}
      \begin{equation}
      {\bf v}_{F}\cdot\frac{\partial}{\partial{\bf r}}\widehat{G}_{\omega}({\bf v}_{F},{\bf r})
      +[\omega\widehat{\tau}_{3}+\widehat{\Delta}({\bf v}_{F},{\bf r}), \widehat{G}_
      {\omega}({\bf v}_{F},{\bf r})]=0\label{eilenberger}
      \end{equation}
      where
      \begin{equation}
     \widehat{\Delta }=\left(
      \begin{array}{cc}
      0 & \Delta \\
      \Delta ^{\dagger } & 0
      \end{array}
      \right) ,\quad \widehat{G}_{\omega }({\bf v}_{F},{\bf r})=\left(
      \begin{array}{cc}
      g_{\omega } & f_{\omega } \\
      f_{\omega }^{\dagger } & -g_{\omega }
      \end{array}
      \right)
      \end{equation}
${\Delta}$ is
the superconducting order parameter, $\widehat{\tau}_{3}$ is the
Pauli matrix, and $\widehat{G}_{\omega }({\bf v}_{F}, {\bf r})$ is
the matrix Green function which depends on the electron velocity
on the Fermi surface ${\bf v}_{F}$, the coordinate ${\bf r}$ and
Matsubara frequency ${\omega}=(2n+1)\pi T$, with $n$ and $T$ being
an integer number and temperature respectively. Also the
normalization condition
\begin{equation}
g_{\omega }={\sqrt{1-f_{\omega }f_{\omega }^{\dagger
}}}\label{norm}
\end{equation}
with $f_{\omega}^{\dagger}$ being time-reversal counterpart of
$f_{\omega}$ should be satisfied by solutions of the Eilenberger
equations. In general, $\Delta $ depends on the direction of ${\bf
v} _{F}$ and ${\bf r}$ and it can be determined by the
self-consistent equation
\begin{equation}
\Delta ({\bf v}_{F}, {\bf r})=2\pi N(0)T\sum\limits_{\omega >0}
\left< V( {\bf v}_{F}, {\bf v}_{F}^{\prime })f_{\omega }({\bf
v}_{F}^{\prime }, {\bf r}) \right>_{{\bf v}_{F}^{^{\prime
}}}\label{self consistent}
\end{equation}
and the current density by
\begin{equation}
 {\bf j(r)}=-4\pi ieN(0)T\sum\limits_{\omega }\left\langle {\bf v}%
 _{F}g_{\omega }({\bf v}_{F}, {\bf r})\right\rangle _{{\bf v}_{F}}
 \label{current}
 \end{equation}
respectively, where $V({\bf v}_{F}, {\bf v}_{F}^{\prime })$ is the
interaction potential, $N(0)$ is the 2D density of states at the
Fermi surface for each spin projection and $\left<... \right>$ is
the averaging over directions of ${\bf v}_{F}$. Solution of the
matrix equation (\ref{eilenberger}) together with self-consistency
gap equation (\ref{self consistent}), Maxwell equation for
superfluid velocity and normalization condition determines the
current ${\bf j(r)}$ in the system. The thickness of $d$-wave
superconductors are assumed to be smaller than coherence length
${\xi}_{0}={\frac{\hbar v_{F}}{\pi \Delta}}$. Thus the spatial
distributions of $\Delta({\bf r})$ and ${\bf j}({\bf r})$ depend
only on the coordinates in the plane of the film and Eilenberger
equations (\ref{eilenberger}) reduce to 2D
equations.\\
Also equations (\ref{eilenberger}) for Green functions
$\widehat{G}_{\omega } ({\bf r}, {\bf v}_{F})$ have to be
supplemented by the condition of specular reflection at the region
$(x=0, |y|\geq a)$ and continuity of solutions across the point
contact $(x=0, |y|\leq a)$. Far from the contact the Green
functions should be coincident with the bulk solutions and the
current should be homogeneous transport current along the
$y$-axis.\\
In this formalism, the current has to be determined by a
self-consistent gap equation (\ref{self consistent}) together with
the Maxwell equation for superfluid velocity (Ampere law $
\frac{d^2 A_y(x)}{d x^2}=-\mu_0 J^{tot}_y(x)$ with $\mu_0$ for
free space and $v_{s}=-\frac{e}{mc}A_{y}(x)$ for superfluid
velocity as in paper \cite{FOG,ANNETT} ). While for simplicity,
the self-consistency of order parameter is ignored and a step
function is considered for spatial dependence and we do not
consider effect of current distribution on the superfluid
velocity. We believe that as in the papers \cite{AMIN1,AMIN2}, the
self-consistent investigation of d-wave Josephson junction does
not show a qualitative different from the
non-self-consistent results.\\
For $\Delta$ and $\bf v_{s}$ being constants at each half plane an
analytical solution for Eilenberger equations can be found by the
method of integration along the quasi-classical trajectories of
quasi-particles. In any point, ${\bf r}=(x,y)$, all ballistic
trajectories can be categorized as transit and
non-transit trajectories.\\
For the transit trajectories the Green functions satisfy
continuity at the contact and the non-transit trajectories satisfy
the specular reflection condition at the partition, $(x=0,|y|\geq
a)$. Also, all transit and non-transit trajectories should satisfy
the boundary conditions in the left and right bulks. Making use of
solutions of the Eilenberger equations, the following expression
is obtained for current at the slit:
\begin{equation}
\hspace{-2.5cm}{\bf j}(x=0,\left\vert y\right\vert <a,\phi ,{\bf
v}_{s},\alpha_l,\alpha_r)=\label{current2}4\pi  e
N(0)v_{F}T\sum\limits_{\omega>0}\left\langle \widehat{{\bf v}}
\frac{\widetilde{\omega}(\Omega_l+\Omega_r)-i\eta\Delta_l\Delta_r\sin
{\phi}}{\Omega_l\Omega_r+\widetilde{\omega}^{2}+\Delta_l\Delta_r\cos\phi}\right\rangle_{\widehat{{\bf
v}}}
\end{equation}
where, $\Delta_{l,r}=\Delta_0\cos(2(\theta-\alpha_{l,r}))$ for
$d_{x^{2}-y^{2}}$ symmetry $\Omega_{l,r} =\sqrt{\widetilde{\omega
}^{2}+\Delta_{l,r} ^{2}}$, $\widetilde{\omega }=\omega +i{\bf
p}_{F}{\bf v}_{s}$ with $\omega$ being Matsubara frequency and
$\bf v_{s}$ is super-fluid velocity, $\widehat{{\bf v}}={\bf
v}_{F}/v_{F}$ is the unit vector, $\eta =sign(v_{x})$. In this
non-stationary Josephson junction, ${\bf v}_{s}\neq0$, the current
has both ${\bf j}_{x}$ and ${\bf j}_{y}$ components. We define the
Josephson current, external transport current, spontaneous current
and interference current as:
\begin{equation}
\hspace{-2.5cm}{\bf j}_{Josephson}={\bf j}(\phi,{\bf
v}_{s},\alpha_l,\alpha_r)_{\hat{x}}
\end{equation}
\begin{equation}
\hspace{-2.5cm}{\bf j}_{Transport}={\bf j}(\phi=0 ,{\bf
v}_{s}\neq0,\alpha_l=\alpha_r=0)_{\hat{y}}={\bf j}_{Bulk},
\end{equation}
\begin{equation}
\hspace{-2.5cm}{\bf j}_{Spontaneous}={\bf j}(\phi\neq0 ,{\bf
v}_{s}=0,\alpha_l\neq0,\alpha_r\neq0)_{\hat{y}}
\end{equation}
\begin{equation}
\hspace{-2.5cm}{\bf j}_{JT}={\bf j}(\phi,{\bf
v}_{s},\alpha_l,\alpha_r)_{\hat{y}}-{\bf j}_{Spontaneous}-{\bf
j}_{Transport}
\end{equation}
respectively. The Josephson current, ${j}_{J}=j_{x}$, is normal to
the interface between superconductors as was considered by B. D.
Josephson in \cite{JOS} and parallel component of current ${\bf
j}_{y}$ consists of three terms of current, an external transport
current, spontaneous current and "interference" current. The new
current term, "interference" current, depends on the super-fluid
velocity, orientations with respect to the interface and phase
difference between order parameters. This term of current, ${\bf
j}_{JT}$, is completely different from the transport
current on the banks ${\bf j}_{T}$.\\
Thus in addition to the spontaneous current for the stationary
$d$-wave Josephson junction that is investigated in Ref.\cite{RAS}
and transport current, we observe another current parallel to the
interface (${\bf j}_{JT}$). In particular, at $\phi\simeq\pi$ it
may go into the opposite direction to the external transport
current on the banks (depending on the orientations). This sign
reversal of tangential supercurrent which is origin of vortex-like
currents near the orifice has been observed already in
Ref.\cite{FSR} for the case of $d-$wave junction. At two sides of
vortex-like currents, currents are flowing parallel and
antiparallel to the external supercurrent at the superconducting
banks. In the Ref.\cite{FSR} a
superconductor-normal-superconductor trajectory for particles has
been considered while there is one superconductor coupled the
normal metal. So, because of one superconductor in the structure
of Ref.\cite{FSR} it was impossible to consider the phase
difference. In the case of junction between $s-$wave
superconductors in Ref.\cite{KOLGHOL}, it was observed that sign
reversal can be seen only for $\phi\simeq\pi$. So, for $s-$wave
counterpart of setup of paper \cite{FSR} it is impossible to see
the sign reversal because phase difference $\phi$ does not meaning
for system including one superconductor. While in our calculations
in this paper we have observed that, this sign reversal and vortex
appearance can be seen at $\phi=0$, depending on mis-orientation.
For the case of $d-$wave in paper of \cite{FSR} in the absence of
phase difference, a sign reversal of current has been observed as
well as our results for suitable mis-orientation of
$ab-$planes($\alpha_l=0, \alpha_r=\frac{\pi}{4}$ or planes
(1,0,0),(1,1,0)). The planes $\alpha_l=0$ and
$\alpha_r=\frac{\pi}{4}$ are corresponding to planes $(100)$ and
$(111)$ respectively.

\begin{figure}[ht]\centering{
\resizebox{\textwidth}{0.5\textheight}{\includegraphics{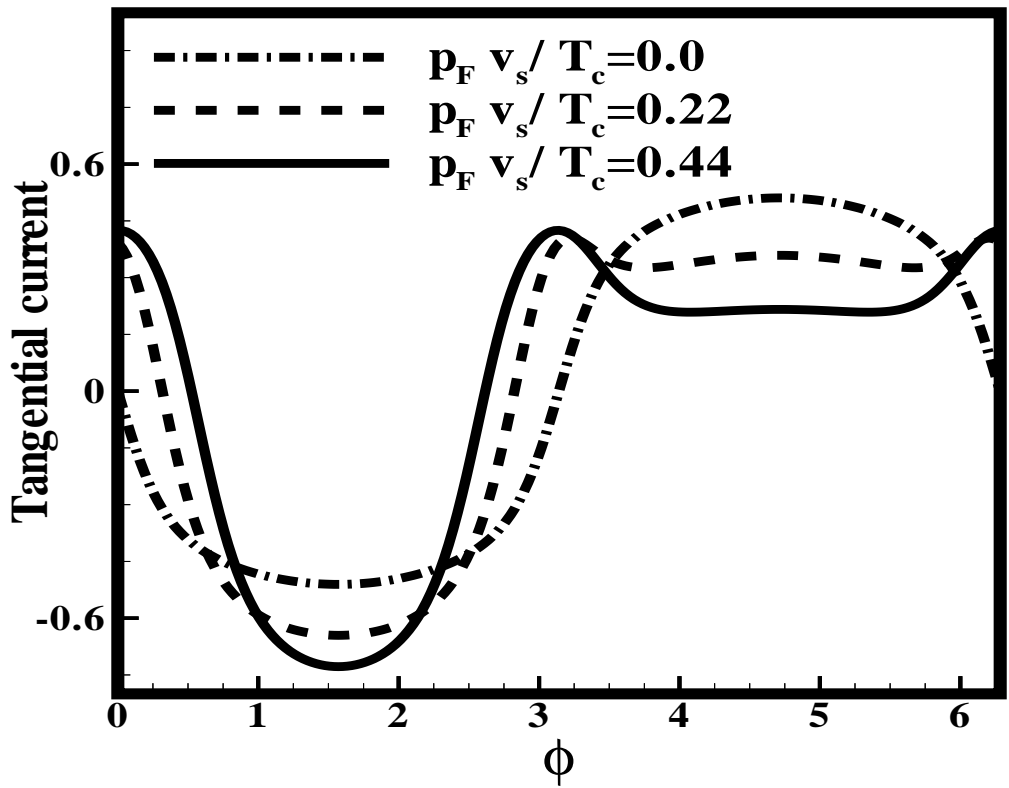}}}\caption{\protect\small
\ Tangential current $j_{y}$ versus phase $\phi $ for $T/T_c=0.1$,
$\alpha_{l}=0$ and $\alpha_{r}=\frac{\pi}{4}$. \hskip0truecm}
\label{fig4}
\end{figure}
\begin{figure}[ht]\centering{
\resizebox{\textwidth}{0.5\textheight}{\includegraphics{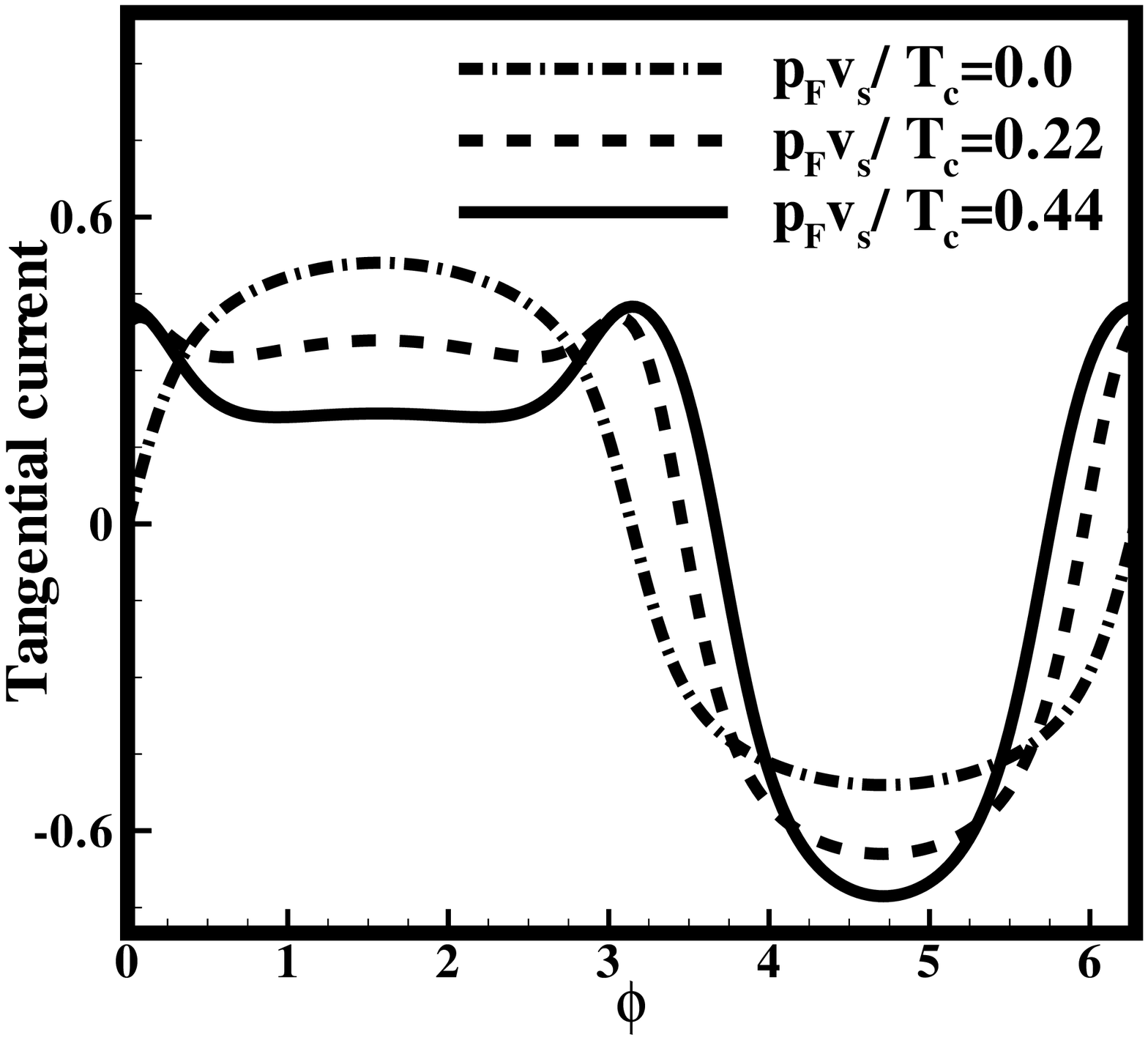}}}
\caption{\protect\small \ Tangential current $j_{y}$ versus phase
$\phi $ for $T/T_c=0.1$, $\alpha_{l}=0$ and
$\alpha_{r}=-\frac{\pi}{4}$. \hskip0truecm} \label{fig3}
\end{figure}
\begin{figure}[ht]\centering{
\resizebox{\textwidth}{0.5\textheight}{\includegraphics{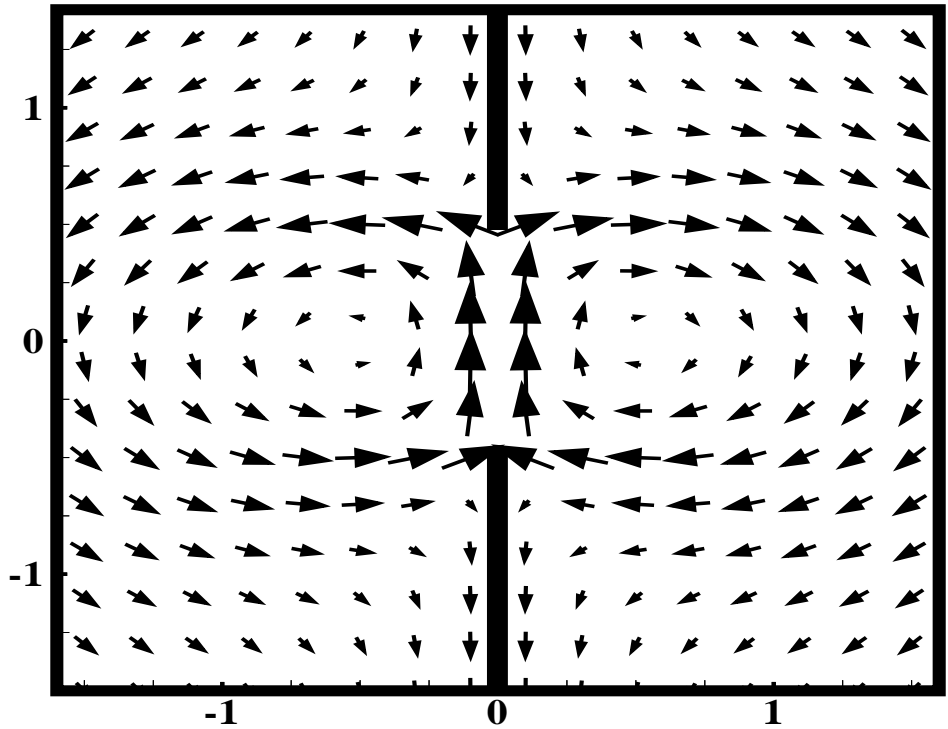}}}
\caption{{\protect\small Vector plot of the current for
$T/T_c=0.1$, $P_Fv_s/\Delta_0(0)=0.5$, $\phi=\pi$,
$\alpha_l=\alpha_r=0$. Axes are marked in the units of $a$.
Because of our non-self-consistent formalism $\xi_{0}= 5a$. It is
similar to the $s$-wave junction in \cite{KOLGHOL}.}
\hskip0truecm} \label{fig7}
\end{figure}
\begin{figure}[ht]\centering{
\resizebox{\textwidth}{0.5\textheight}{\includegraphics{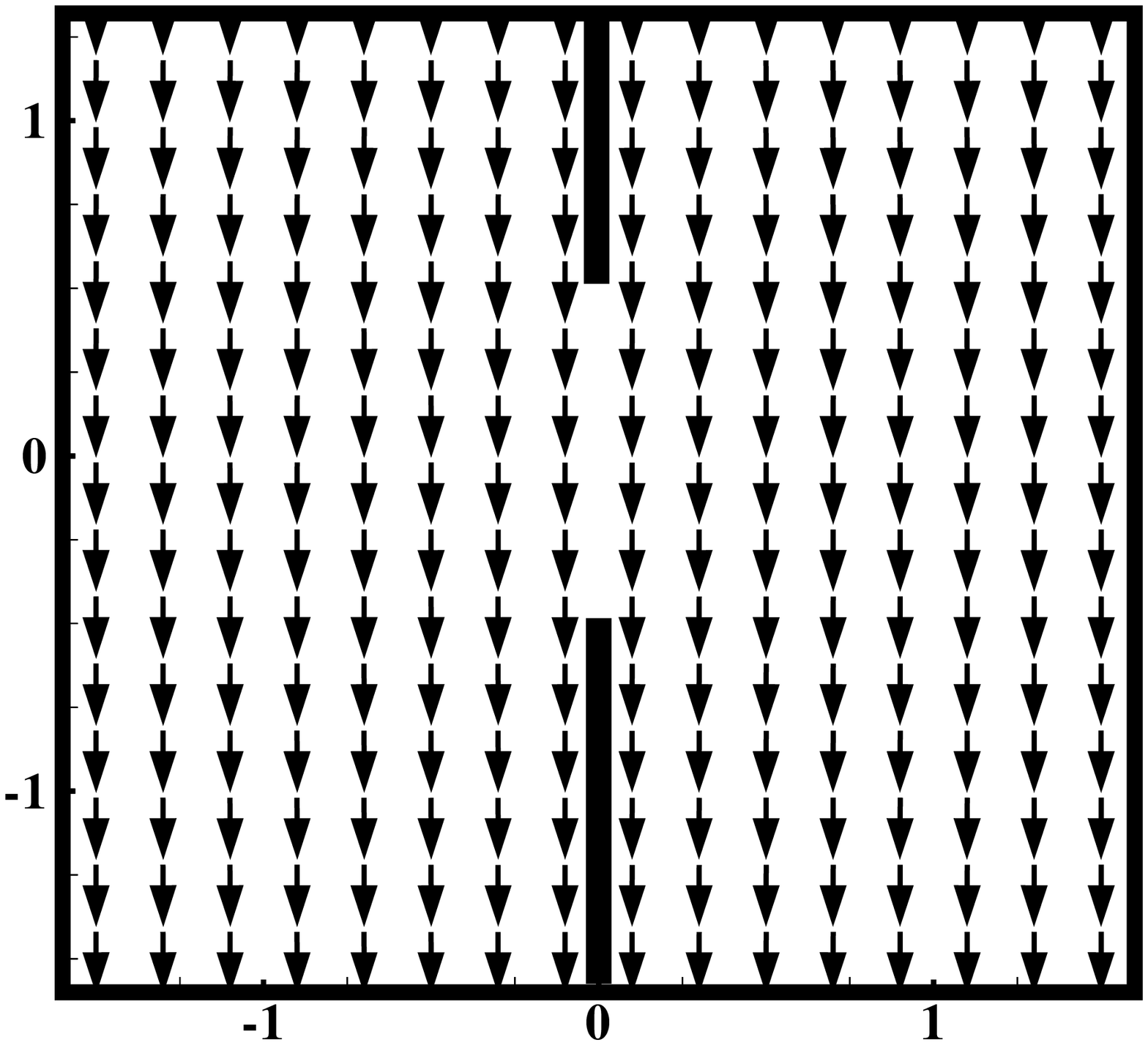}}}
\caption{{\protect\small Vector plot of the current for $\phi=0$,
$\alpha_l=0$, $\alpha_r=0$ and $T/T_c=0.1$,
$P_Fv_s/\Delta_0(0)=0.5$. Vortices disappeared but transport
supercurrent flows.} \hskip0truecm} \label{fig9}
\end{figure}
\section{discussion}\label{3}
\begin{figure}[ht]\centering{
\resizebox{\textwidth}{0.5\textheight}{\includegraphics{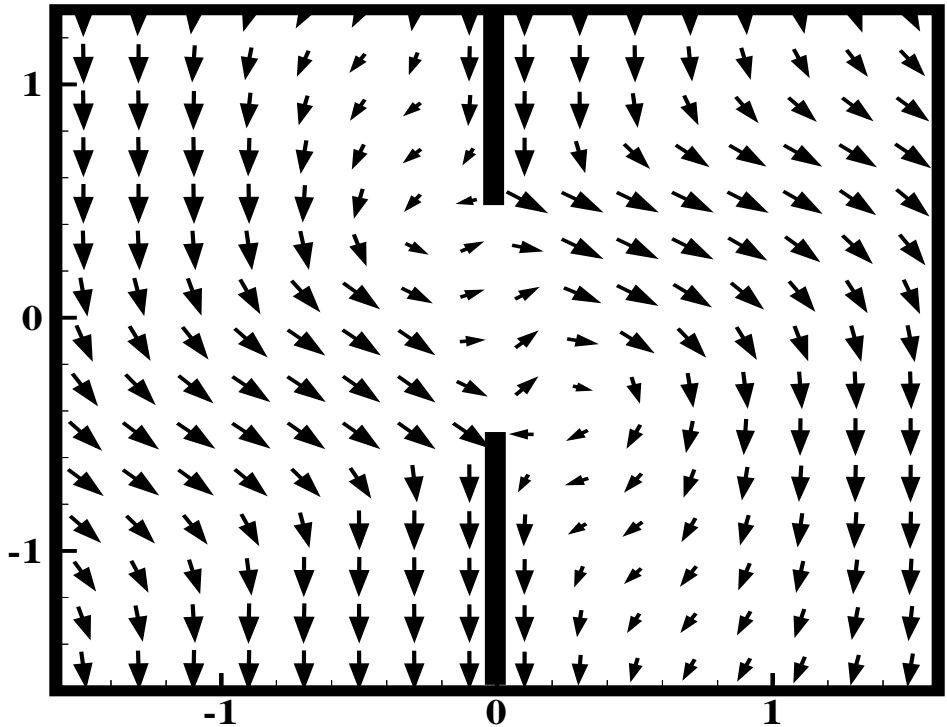}}}\caption{{\protect\small
Vector plot of the current for $\phi=\pi$, $\alpha_{l}=0$,
$\alpha_{r}=\frac{\pi}{4}$ and $T/T_c=0.1$,
$P_Fv_s/\Delta_0(0)=0.5$. Axes are marked in the units of $a$.} }
\label{fig6}
\end{figure}

\begin{figure}[ht]\centering{
\resizebox{\textwidth}{0.5\textheight}{\includegraphics{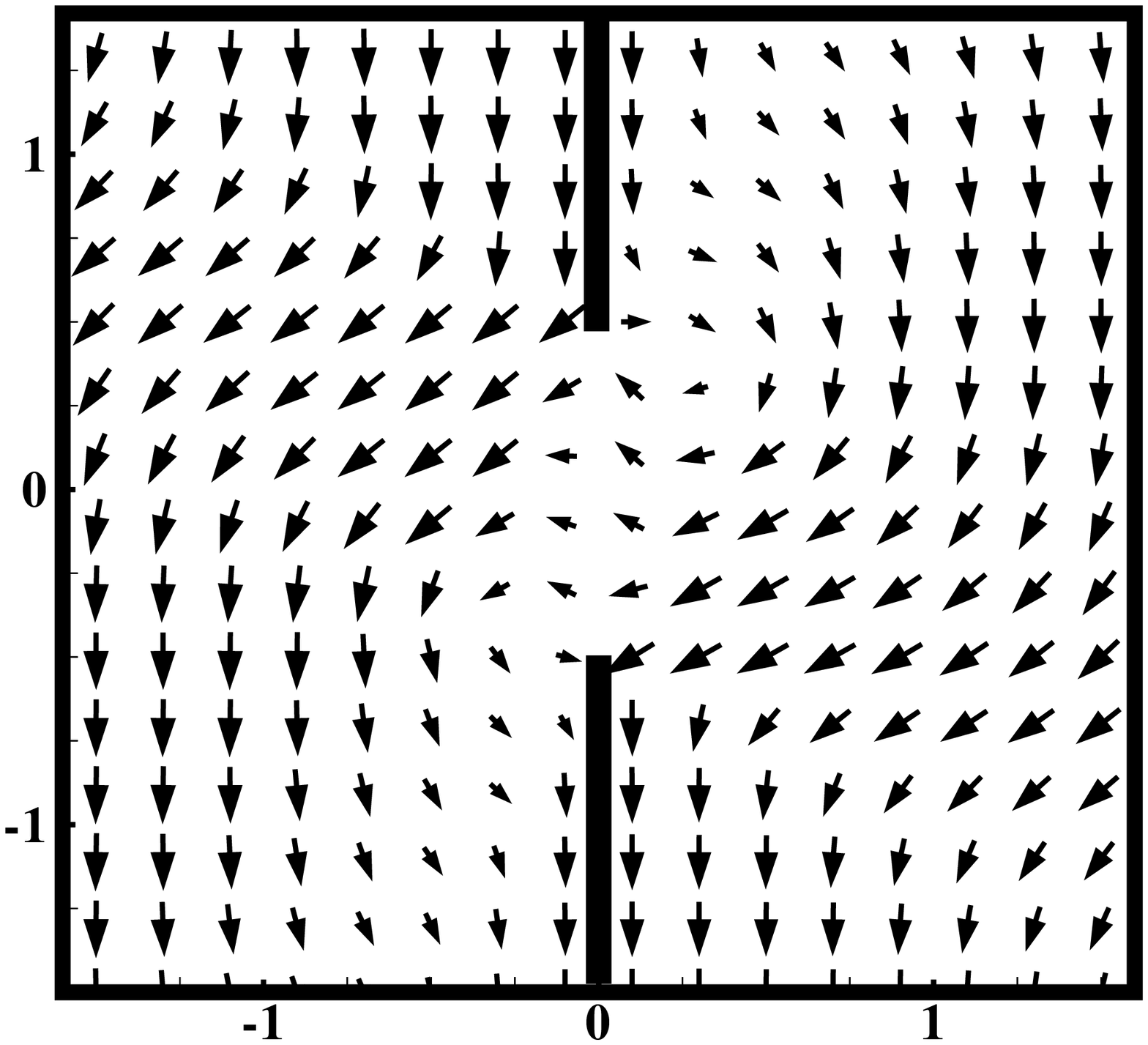}}}\caption{{\protect\small
Vector plot of the current for $\phi=\pi$, $\alpha_{l}=0$,
$\alpha_{r}=-\frac{\pi}{4}$ and $T/T_c=0.1$,
$P_Fv_s/\Delta_0(0)=0.5$. The axis of vortices is rotated.}
\hskip0truecm} \label{fig8}
\end{figure}

The effects of mis-orientation and super-fluid velocity on the
current-phase relation and current distribution are investigated
numerically and the results are following.\\
1) The vortex-like currents appear at values of $\phi$ when the
parallel current is maximum and positive while the external
transport
current is negative (Figs.\ref{fig2},\ref{fig5},\ref{fig4} and \ref{fig3}).\\
2) At $\alpha_l=\alpha_r=0$ and arbitrary $\phi$, the current
distributions and current-phase graphs are identical with the
$s$-wave which is investigated in Ref.\cite{KOLGHOL}. For example
at $\alpha_l=\alpha_r=0$ and $\phi=\pi$ two anti-symmetric
vortex-like currents are observed and their common
axis is normal to the interface (Fig.\ref{fig7}).\\
3) At $\alpha_{l}=0$, $\alpha_{r}=\pm\frac{\pi}{4}$ and
$\phi=\pi$, two vortex-like currents are observed. Their common
axis is rotated as much as $\mp\frac{\pi}{4}$ (Fig.\ref{fig6} and
Fig.\ref{fig8}). This case occurs for $\alpha_{l}=0$, $\phi=\pi$
and arbitrary $\alpha_{r}$ with the
rotated axis rotating as much as $(-\alpha_{r})$.\\
4) The appearance of vortex-like currents can be controlled by
mis-orientation. For example, at $\phi=0$ and
$\alpha_l=\alpha_r=0$, we cannot observe the vortex-like currents
as we cannot observe in $s$-wave junction (Fig.\ref{fig9}).
However, for $\phi=0$, $\alpha_{l,r}=0$,
$\alpha_{r,l}=\frac{\pi}{2}$ the vortex-like currents appear since
from (\ref{current2}) and for $d_{x^2-y^2}$ symmetry we have :
\begin{equation}
{\bf j}({\bf r},\phi=\pi ,{\bf v}_{s},\alpha_l,\alpha_r)={\bf
j}({\bf r },\phi=0 ,{\bf
v}_{s},\alpha_l,\alpha_r+\frac{\pi}{2}).\label{symmetry}
\end{equation}
5) In Figs.\ref{fig5},\ref{fig4},\ref{fig3} it is observed that
for $\phi=0$ and consequently without any external magnetic flux,
the interference between coherent current-states can occur and the
vortex-like currents can be observed because mis-orientation plays
the role of the phase difference in Eq.(\ref{symmetry}).\\
6) The parallel current, ${\bf j}_y$, is plotted in terms of the
phase difference for different super-fluid velocities and at
$\alpha_l=\alpha_r=0$, the maximum value of current and appearance
of vortex-like currents occur at $\phi=\pi$. In this case, far
from $\phi=\pi$, we observe a constant current that is the external transport current on the banks (Fig.\ref{fig2}).\\
7) For $\alpha_{l,r}=0$, $\alpha_{r,l}=\pm\frac{\pi}{4}$ the
maximum values of the parallel current, ${\bf j}_y$, and
consequently
the vortices appear at $\phi=0$, $\phi=\pi$, $\phi=2\pi$ (Figs.\ref{fig4},\ref{fig3}).\\
8) For $\alpha_{l,r}=0$, $\alpha_{r,l}=\frac{\pi}{2}$ the
current-phase graphs are similar to the $\alpha_l=\alpha_r=0$ but
a displacement as much as $\pi$ occurs
(Figs.\ref{fig2},\ref{fig5}). Thus the vortices can be observed at
$\phi=0$ and $\phi=2\pi$. But at $\phi=\pi$, we do not observe the
vortex-like currents. This can be another difference between
conventional and unconventional Josephson junctions.\\
9) The superposition of dash lines in Figs.\ref{fig4},\ref{fig3}
for zero super-fluid velocity, and all the lines of
Fig.\ref{fig2}, for zero mis-orientations gives us the three lines
of Figs.\ref{fig4},\ref{fig3} apparently. This means that in this
case the tangential current is algebraic sum of transport current,
"interference" current\cite{KOLGHOL} and spontaneous current \cite{RAS}.\\
10) The tangential current for $\alpha_l=\alpha_r=0$ and
$\alpha_{l,r}=0,\alpha_{r,l}=\frac{\pi}{2}$ is an even function of
Josephson phase $\phi$ but for
$\alpha_{l,r}=0,\alpha_{r,l}=\pm\frac{\pi}{4}$ it is neither even
nor odd function of phase difference $\phi$ and the symmetry will be broken.\\
The superfluid of pairs creates the transport current but the
spontaneous current is produced by mis-orientation plus to the
phase difference. However the "interference" current depends on
all of the parameters (phase difference, mis-orientation and
superfluid velocity) and it can
be the result of the non-locality of supercurrent.\\
The spatial distributions of the order parameter and the current
near and precisely at the contact are calculated using the Green
function along the transit and non-transit trajectories
numerically. Transit trajectories for each point are coming from
the orifice (transparent part of interface $x=0, |y|\leq a$) while
non-transit trajectories form the remainder part of interface
which is impenetrable (reflective part of interface $x=0, |y|\geq
a$).
The current distributions are calculated and simulated numerically
for $T/T_c=0.1$, $p_Fv_s/\Delta_0(0)=0.5$ and different choices of
mis-orientation for $\phi=\pi$
(Figs.\ref{fig6},\ref{fig7},\ref{fig8}) and $\phi=0$
(Fig.\ref{fig9}). Following the Ref.\cite{KOLGHOL}, the
"interference" current that plays a central role for production of
vortex-like current is found. At the phase values
$0<\phi<\frac{\pi}{2}$, $\frac{3}{2}\pi<\phi<2\pi$ and
$\alpha_l=\alpha_r=0$, the interference current is very small thus
the total current is close to the vector sum of transport,
spontaneous and Josephson currents. But, for
$\frac{\pi}{2}<\phi<\frac{3}{2}\pi$ the "interference" current
appears and the total current deviates from the vector sum of the
three old currents \cite{KOLGHOL}. Also, for
zero mis-orientation the spontaneous current is zero.\\
The "interference" current is always anti-parallel to the
transport current. But the spontaneous current may be parallel or
anti-parallel to the transport current. If algebraic sum of
transport, "interference" and spontaneous current is anti-parallel
to the transport current on the banks, we can find  the vortex-like currents (Fig.\ref{fig7}).\\
Thus the appearance of the vortex-like currents can be controlled
by mis-orientation and phase difference. It is remarkable that,
far from the contact $|{\bf r}|\sim {\xi_{0}}> a$ for all $\phi$s,
mis-orientations, temperatures $T$ and superfluid velocities, the
distributions of currents tend to the tangential transport
currents on the banks.\\
\section{near the critical temperature}\label{4}
For temperatures close to the critical temperature, $T_{c}-T \ll
T_{c} $  problem is solvable pure analytically \cite{KOLGHOL}. At
the contact we have
\begin{equation}
{\bf j}={\bf j}_{J}+{\bf j}_{spont}+{\bf j}_{T}+{\bf
j}_{JT}\label{sepertc}
\end{equation}
\begin{equation}
{\bf j}_{J}=2j_{c}\sin \phi \left\langle \widehat{{\bf
v}}_{x}sign(v_{x})(\frac{\Delta_l\Delta_r}{\Delta_0^{2}})\right\rangle
_{\widehat{{\bf v}}}\label{josephtc}
\end{equation}
\begin{equation}
{\bf j}_{spont}=2j_{c}\sin \phi \left\langle \widehat{{\bf
v}}_{y}sign(v_{x})(\frac{\Delta_l\Delta_r}{\Delta_0^{2}})\right\rangle
_{\widehat{{\bf v}}}\label{sponttc}
\end{equation}
\begin{equation}
{\bf j}_{T}=-j_{c}k\left\langle \widehat{{\bf v}}\widehat{v}%
_{y}(\frac{\Delta_l\Delta_r}{\Delta_0^{2}})\right\rangle
_{\widehat{{\bf v}}}\label{transtc}
\end{equation}
\begin{equation}
{\bf j}_{JT}=j_{c}k(1-\cos \phi )\left\langle\widehat{{\bf
v}}\widehat{v}_{y}(\frac{\Delta_l\Delta_r}{\Delta_0^{2}})\right\rangle
 _{\widehat{{\bf v}}}\label{intertc}
\end{equation}
where, as in Ref.\cite{KOLGHOL}
\begin{equation}
j_{c}\left( T,v_{s}\right) =\frac{\pi
|e|N(0)v_{F}}{8}\frac{{\Delta_0}^{2}\left( T,v_{s}\right)
}{T_{c}}\label{jc}
\end{equation}
is a critical current of the contact at $(T_{c}-T)\ll T_{c}$, $k$
is a standard notation
\begin{equation}
k=(14\varsigma(3)/\pi^{3})(v_{s}p_{F}/T_{c}).
\end{equation}
For the high value of temperatures near the $T_c$, the critical
values of currents have a linear dependence on the
${\Delta_0}^{2}$, which can be replaced by
$\Delta_0=\sqrt{\left(\frac{32\pi^2}{21\zeta(3)}\right)T_c(T_c-T)}.$
On the other hand the spontaneous and Josephson currents are
sinusoidal function of phase difference as is expected for the
currents near the $T_c$, in spite of these two terms of current
the "interference" current is an even function of the phase
difference. The current is divided into the four parts, Josephson
current ${\bf j}_{J}$, spontaneous current, transport current in
the banks ${\bf j}_{T}$ and the "interference" current ${\bf
j}_{JT}$. It is observed that, the currents generally and near the
$T_c$ obviously, depend not only on the mis-orientation
$|\alpha_l-\alpha_r|$ but also depend on the orientations with
respect to the interface. Because in the expressions
$\left\langle\widehat{{\bf v}}\Delta_l\Delta_r...\right\rangle$,
the result of angular integrations may include both
$|\alpha_l-\alpha_r|$ and $|\alpha_l+\alpha_r|$ terms. For
example, for Josephson and spontaneous currents in
(\ref{josephtc}) and (\ref{sponttc}) by angular integration on the
Fermi surface we have
\begin{equation}
{\bf j}_{J}=(\frac{2j_{c}\sin
\phi}{15\pi})[15\cos(2\alpha_l-2\alpha_r)-\cos(2\alpha_l+2\alpha_r)]\widehat{{\bf
x}}\label{integrated1}
\end{equation}
and for spontaneous current
\begin{equation}
{\bf j}_{spont}=(\frac{-8j_{c}\sin
\phi}{15\pi})\sin(2\alpha_l+2\alpha_r)\widehat{{\bf
y}}\label{integrated2}
\end{equation}
respectively. Also it is obtained that for $\phi=\pi$ and exactly
at the contact, the "interference" current ${\bf j}_{JT}$ is
anti-parallel to the ${\bf j}_{T}$. For $\phi=\pi$ the
"interference", Josephson and spontaneous currents are ${\bf
j}_{JT}=-2{\bf j}_{T}$, ${\bf j}_J=0$ and ${\bf j}_{spont}=0$
respectively, consequently it is obtained that ${\bf j}_{y}={\bf
j}_{T}+{\bf j}_{JT}+{\bf j}_{spont}=-{\bf j}_{T}.$ In this case
while the Josephson current is zero the terms ${\bf j}_{y}$ and
${\bf j}_{T}$ that are directed opposite to the each other,
control the appearance of vortex-like currents in vicinity of the
point contact. In addition, for $\phi=0$, $\alpha_{l,r}=0$ and
$\alpha_{r,l}=\frac{\pi}{2}$, we can observe the vortex-like
currents. This property can be a difference between $d$-wave and
$s$-wave Josephson junctions. Because in the $s$-wave Josephson
junction the vortex-like currents are observed only for $\phi=\pi$
(Ref.\cite{KOLGHOL}), while in the present calculations for
$d$-wave Josephson junction, the vortex-like currents may appear
even for $\phi=0$. Thus, for fixed values of the temperature and
superfluid velocity, the presence of vortex-like currents can be
controlled
by mis-orientation and phase difference. \\
 \section{Conclusions}\label{5}
The vortex-like currents in vicinity of the point contact are
observed for $d$-wave Josephson junction as well as for $s$-wave
junction in Ref.\cite{KOLGHOL}. The interference current as a
result of non-local supercurrent states, appears. It may flow
opposite to the external transport current. For $\phi=\pi$,
$\alpha_l=0$ and $\alpha_r=\pm\frac{\pi}{4}$ the vortex-like
currents with the rotated axis are observed. But as is obtained in
Ref.\cite{KOLGHOL}, the axis of vortex-like currents in the
$s$-wave Josephson junction is normal to the interface. Thus, this
rotated axis can be used to distinguish between $s$-wave and
$d$-wave junction. Also it can be exerted to distinguish between
the junction between two pure $s$-wave and mixing of conventional
and unconventional order parameters (eg.$d+is$). In addition to
the ZES, this rotated axis can be another "fingerprint" of
$d$-wave pairing symmetry. Another interesting result is, the
behavior of system in the absence of the phase difference. For
$s$-wave system only in the presence of the phase difference,
$\phi\simeq\pi\neq 0$, the vortex-like currents appear
\cite{KOLGHOL}, while for $d$-wave Josephson junction at zero
phase and zero external magnetic flux, it is possible to observe
the vortex-like currents for some mis-orientations. This can be a
theoretical reason that, the mis-orientation plays role instead of
the phase difference (Josephson phase). In the stationary
Josephson junction $\bf v_{s}=0$ the tangential interference
current (spontaneous current) will be observed only for $\phi \neq
0$ but in the opposite case $\bf v_{s}\neq 0$ the tangential
current even in the absence of phase difference may be observed.
In addition, this tangential current can flow in the opposite
direction to the external transport current and this factor can
produce the vortex-like currents. Finally, playing the role of
magnetic Josephson phase ($\phi=\frac{q\Phi}{\hbar}$ where $q$ and
$\Phi$ are electric charge and magnetic flux respectively) by
mis-orientation of superconducting $ab$-planes (pairing symmetry
in the momentum space) is a reason for magnetic nature of pairing
mechanism in the high $T_c$ superconductors which remains as an
unknown and famous problem.

\end{document}